\documentclass[aps,prd,superscriptaddress,showpacs,preprintnumbers]{revtex4}
\usepackage{amssymb, graphics, setspace}
\usepackage{pgfplots}
\usepackage{textcomp}
\usepackage[caption=false]{subfig}

\usepackage{amsmath}
\usepackage{commath}

\usepackage{graphicx,bm}
\usepackage{url}
\usepackage{float}
\usepackage{textcomp}
\usepackage{verbatim}
\begin{document}

\newcommand{\dlt}{\bigtriangleup}
\newcommand{\beq}{\begin{equation}}
\newcommand{\eeq}[1]{\label{#1} \end{equation}}
\newcommand{\insertplot}[1]{\centerline{\psfig{figure={#1},width=14.5cm}}}

\parskip=0.3cm

%\begin{titlepage}

\title{Structures in the  diffraction cone: the "break" and "dip" in high-energy proton-proton scattering}

%\author{X}
%\affiliation{...}

\author{L\'aszl\'o Jenkovszky}
\affiliation{Bogolyubov Institute for Theoretical Physics (BITP),
Ukrainian National Academy of Sciences \\14-b, Metrologicheskaya str.,
Kiev, 03680, UKRAINE; jenk@bitp.kiev.ua}

\author{Istv\'an Szanyi}
\affiliation{Uzhgorod National University, \\14, Universytets'ka str.,  
Uzhgorod, 88000, UKRAINE; sz.istvan03@gmail.com}

\begin{abstract}
Anticipating forthcoming publication by the TOTEM collaboration of low-$|t|$ elastic scattering data at $\sqrt{s}=13$ and $2.76$ TeV, hereby we emphasize the correlation between two prominent structures seen upon the otherwise exponential diffraction cone, namely 
a "break" staying fixed around $t \approx -0.1$ GeV$^2$ and a dip moving  with energy logarithmically inwards; while at the ISR the two structures are separated by a distance of about $1$~GeV$^2$, at the LHC the dip comes close to the periphery of the "break", thus affecting its parametrization. An unbiased disentangling and identification of the break at the LHC should account for this correlation.   

\end{abstract}

\pacs{13.75, 13.85.-t}

\maketitle
\section{Introduction} \label{s1}	
The diffraction cone of high-energy elastic hadron scattering deviates from a purely exponential due to two structures clearly visible in proton-proton scattering, namely a "break" (in fact, a smooth concave curvature) near $t=-0.1$~GeV$^2$, whose position is independent of energy, and the prominent "dip" - diffraction minimum moving slowly (logarithmically) with $s$ towards smaller values of $|t|$, where $s$ and $t$ are the  Mandelstam variables. While in the ISR energy region, $23.5\leq \sqrt{s}\leq 62.5$ GeV, the dip is known to be located near $-t=1.4$ GeV$^2$, at the LHC, $\sqrt {s}=7$ TeV it was found \cite{TOTEMdip} near $t\approx-0.5$ GeV$^2$. Physics of the two phenomena is quite different and still disputable: while the "break" may be related \cite{LNC} to the two-pion threshold required by $t$-channel unitarity, the dip (diffraction minimum) is a consequence of $s$-channel unitarity or absorption corrections to the scattering amplitude. 

\begin{figure}[h] 
	\centering
	\subfloat[TOTEM's 7 TeV measuremnets. The figure is taken from Ref.~\cite{TOTEM7}. The references {[1]} and {[2]} in the legend correspond to our references 1 and 4. \label{Fig:2a}]{%
		\includegraphics[width=0.7\textwidth]{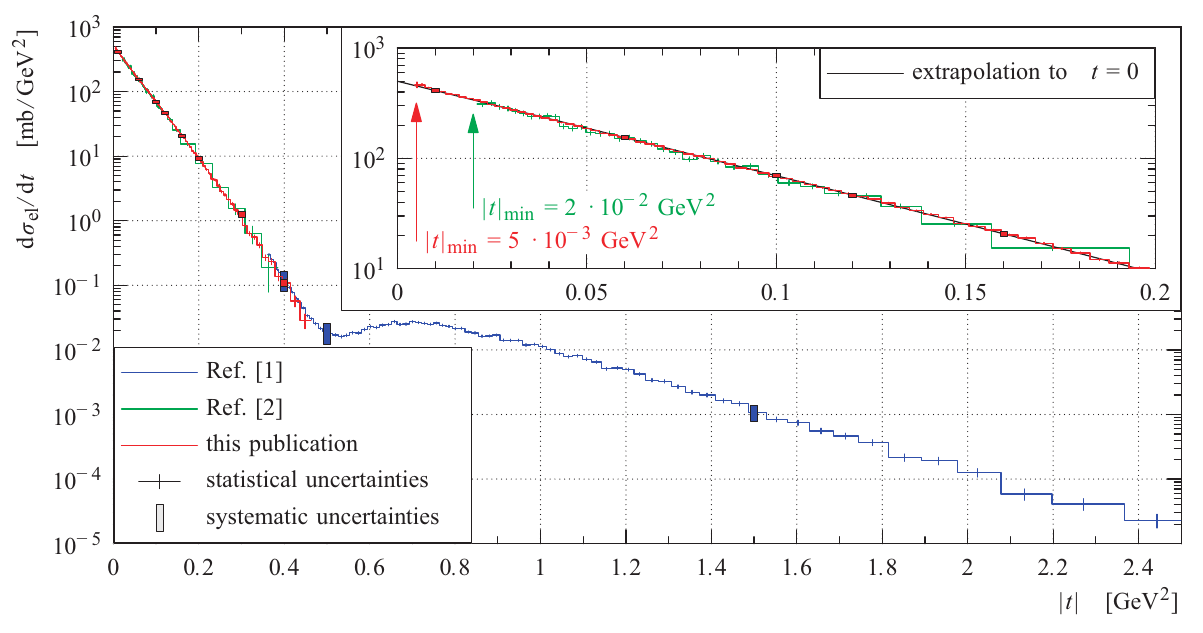}%
	}\quad
	\subfloat[TOTEM's 8 TeV measurements in normalized form  \cite{TOTEM8}. \label{Fig:2b}]{%
		\includegraphics[width=0.7\textwidth]{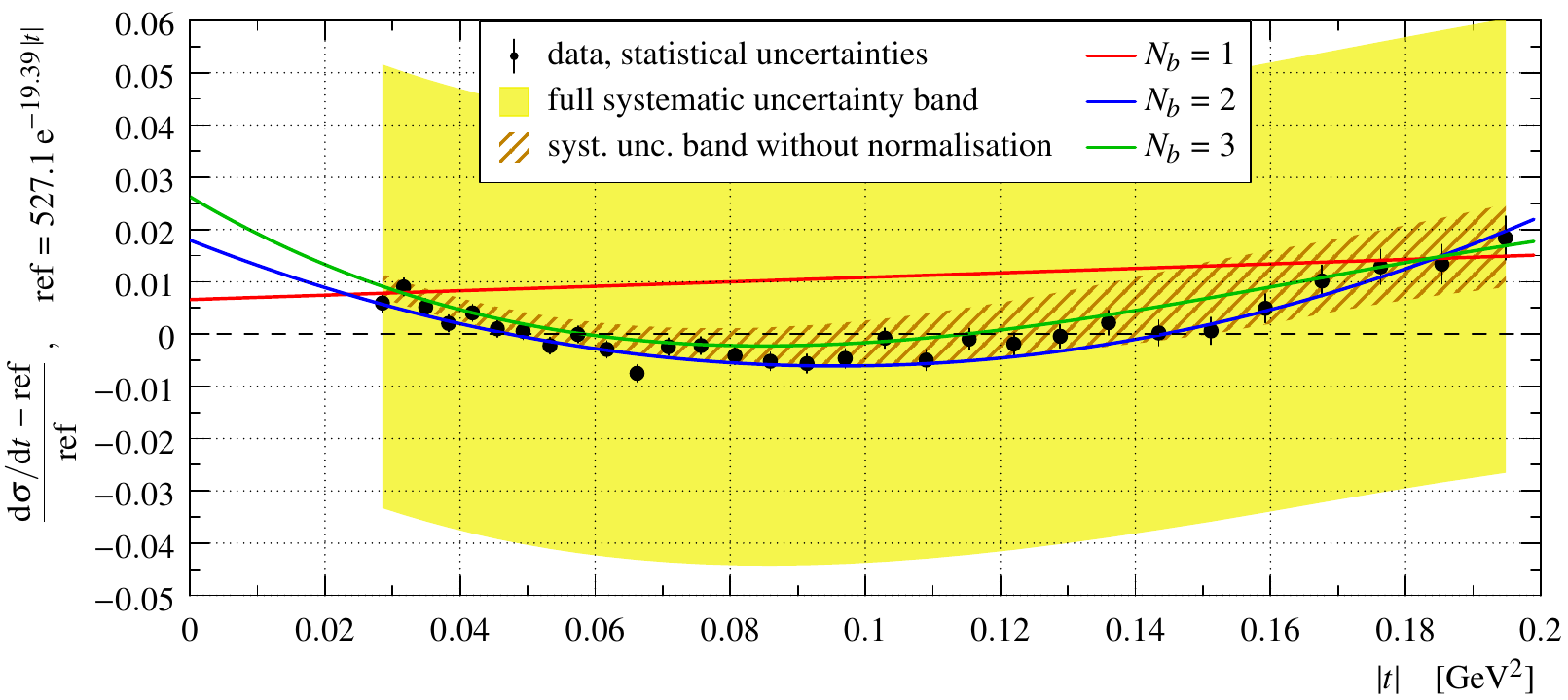}%
	}
	\caption{Results of the TOTEM measurements.}
	\label{Fig:2}
\end{figure}

\begin{figure}[H] 
	\centering
	\subfloat[\label{Fig:1a}]{%
		\includegraphics[width=0.35\textwidth]{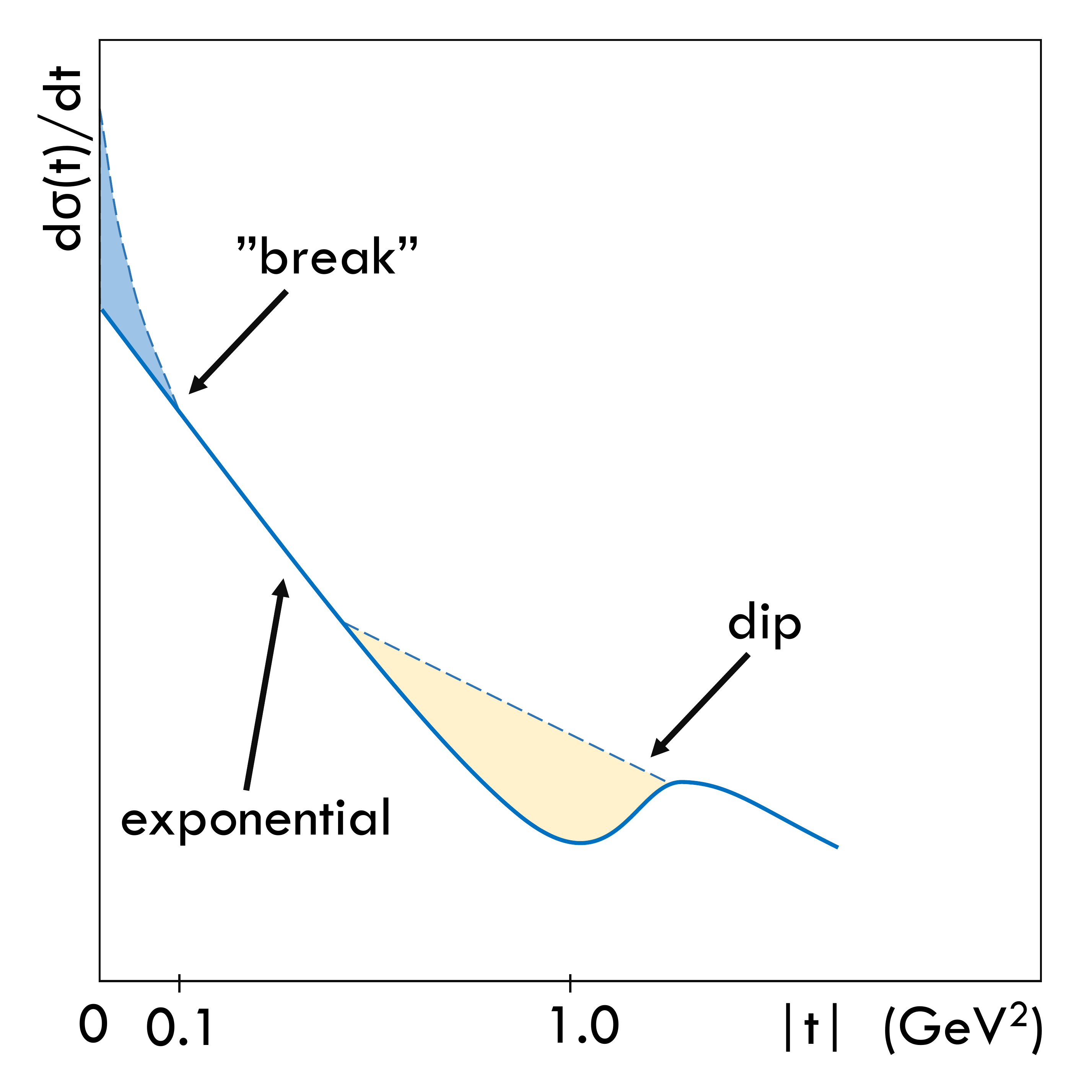}%
	}\quad
	\subfloat[\label{Fig:1b}]{%
		\includegraphics[width=0.35\textwidth]{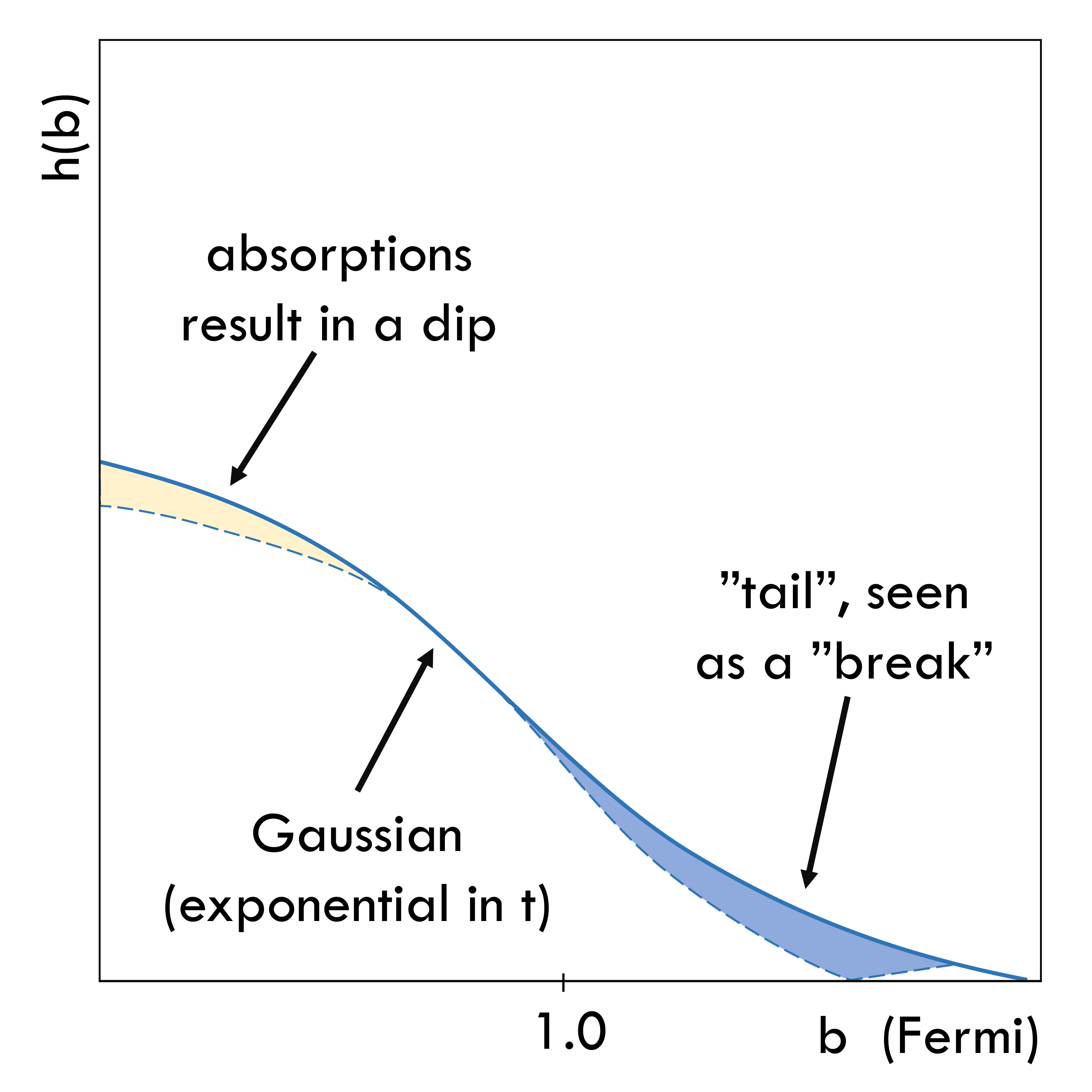}%
	}
	\caption{Schematic (qualitative) view of the "break", followed by the diffraction minimum ("dip"), shown both as function in $t$ and its Fourier transform (impact parameter representation), in $b$. While the "break" reflects the presence of the pion "atmosphere" (clouding) around the nucleon at peripheral values of $b$, the dip results from absorption corrections, suppressing the impact parameter amplitude at small $b$.}
	\label{Fig:1}
\end{figure}

The results of the low-$|t|$ measurements of the  $pp$ cross differential sections at the LHC were  published in Refs. \cite{TOTEM7} and \cite{TOTEM8}. Fig.~\ref{Fig:2} compiles the results of both measurements. While paper \cite{TOTEM7} claims structureless behavior in the cone, the subsequent papers \cite{TOTEM8}, reporting on measurements at $8$ TeV at still smaller values of $|t|$ Fig.~\ref{Fig:1a} clearly show deviation from the linear exponential with a significance grater then 7$\sigma$. Recently, preliminary results at $13$ TeV were made public \cite{TOTEM13}.

\begin{figure}[h] 
	\centering
	\includegraphics[width=.7\textwidth]{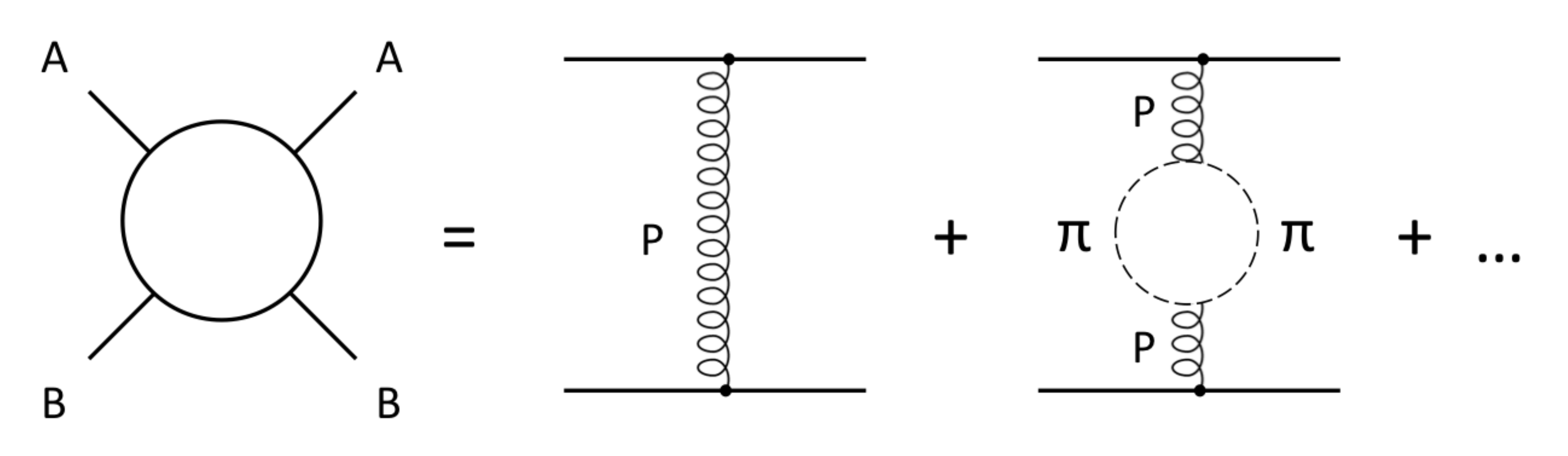}
	\caption{Feynman diagram for elastic scattering with $t$-channel exchange containing a branch point at $t=4m_{\pi}^2$.} 
	\label{Fig:Diagram}
\end{figure}

As illustrated in Fig. \ref{Fig:1}, the "break" corresponds to the nucleon "atmosphere" (pion clouding). The effect, first observed at the ISR was interpreted as manifestation of $t$-channel unitarity, generating a two-pion loop in the cross channel, Fig. \ref{Fig:Diagram}, called by Bronzan \cite{Bronzan} fine structure of the Pomeron. Subsequently it was scrutinized in a number of papers \cite{LNC, BPM, Brazil, RPM}.   

\begin{figure}[H] 
	\centering
	\includegraphics[width=.6\textwidth]{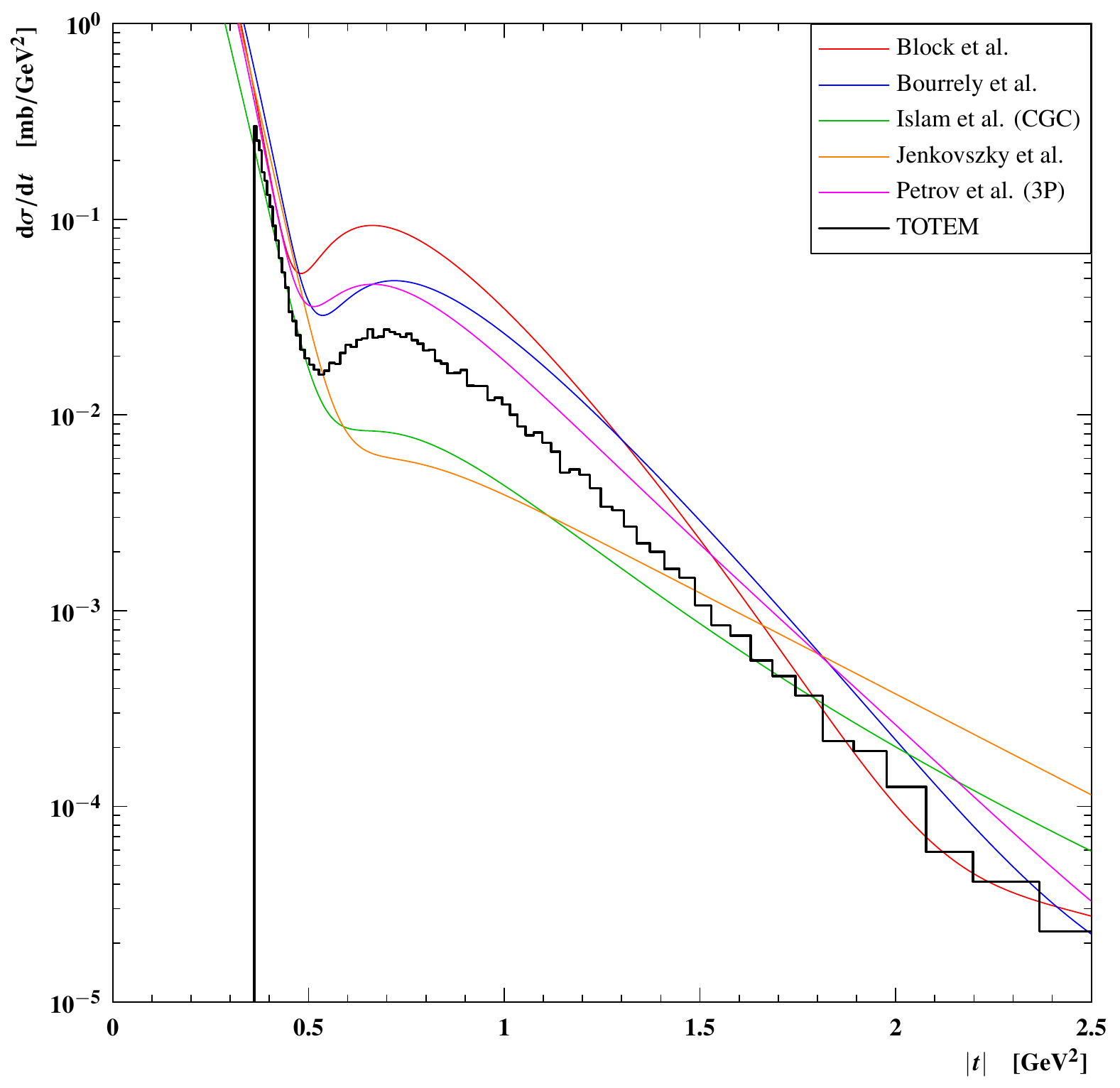}
	\caption{The "dip" in $d\sigma/dt$ at 7 TeV  as seen by \cite{TOTEMdip}, compared to  several model predictions.} 
	\label{Fig:7models}
\end{figure}

The diffraction minimum instead is a consequence of $s$-channel unitarity (or absorption corrections) damping the impact-parameter amplitude at small-$b$, as shown in Fig.~\ref{Fig:1b}. The details are model-dependent therefore the predictions are far from unique, see Fig.~\ref{Fig:7models}. Comprehensive discussions of these phenomena and relevant references may be found {\it e.g.} in Ref. \cite{Reviews}.

In the present paper we do not stick to a particular model of either the "break" or "dip". Instead we wish to attract the experimentalists attention to the possible correlation between the two phenomena at the LHC. To see clearly the the "break", it is necessary to separate it from the influence of the neighbouring dip. It is obvious that better statistics requires more data point, which is equivalent to the extension of the $t$-range considered towards larger values of $|t|$, risking influence of the near-by "dip". The details of the "break" can be identified by optimizing the $t$-range studied/fitted with account of the correlations with the neighbouring dip.

The structures we are discussing are universal for any high-energy hadron scattering, including diffraction dissociation, although they were clearly seen only in proton-proton scattering (at the ISR and at the LHC). Proton-antiproton scattering would be another interested process, however lack Roman Pots in low-$|t|$ measurements of elastic $p\bar p$ scattering at the Tevatron preventing high-precision measurements indispensable for the detection of the tiny "break". Moreover, the dip in $p\bar p$ at the Tevatron is only "rudimental", seen as a shoulder, probably due to the Odderon filling the dip. On the other hand, the structures established in proton-proton scattering are expected to be present also in other hadronic reactions, first of all in $K^+p$ and in diffraction dissociation.

\section{The "break" at the ISR and at the LHC} \label{s2}
In a recent paper \cite{RPM} the low-$|t|$ elastic $pp$ data, including the "break", was scrutinized at several energies. To answer the question about the universality of the fine structure of the diffraction cone (of the Pomeron?!) we have extrapolated the low-$|t|$ structure from the ISR to the LHC. To do so, a Regge-pole model was used, in which the "break" was parametrized by means of a non-linear Pomeron trajectory with a two-pion threshold corresponding to the loop of Fig.~\ref{Fig:Diagram}. 

\begin{figure}[H] 
	\centering
	\subfloat[$\sqrt{s}$=21 and 31 GeV]{%
		\includegraphics[width=0.45\textwidth]{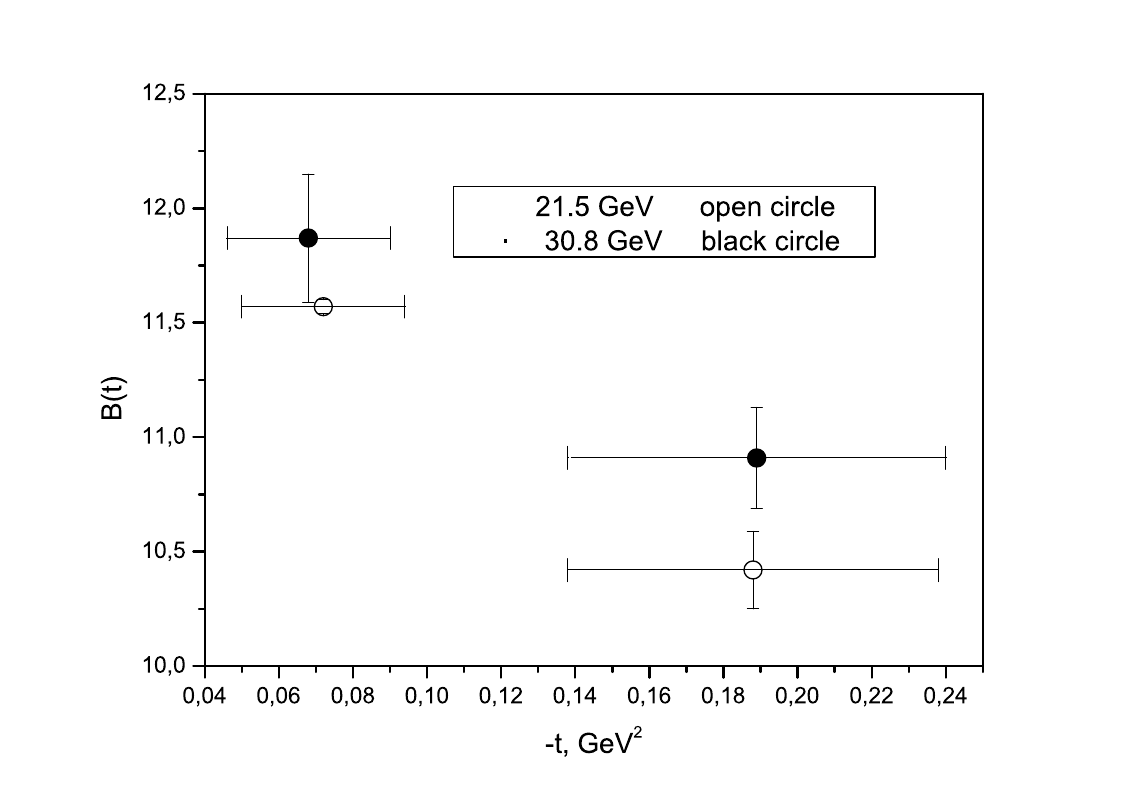}%
	}\quad
	\subfloat[$\sqrt{s}$=45 and 53 GeV]{%
		\includegraphics[width=0.45\textwidth]{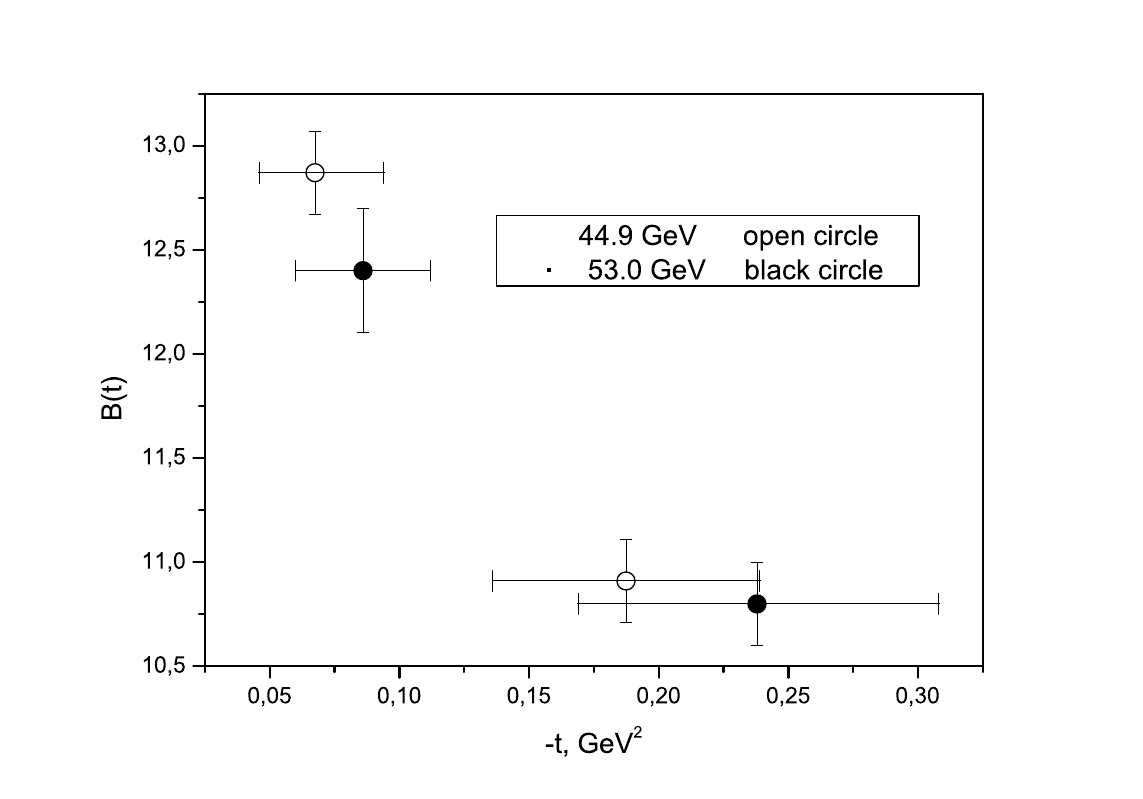}%
	}
	\caption{Local slopes $B(t)$ calculated for the ISR data \cite{Bar}.}
	\label{Fig:slopes}
\end{figure}

The local slope is calculated as $B(s,t)=\frac{d}{dt}ln\frac{d\sigma}{dt}.$ Fig.~\ref{Fig:slopes} shows the slopes calculated from the ISR data of Ref. \cite{Bar}. The slope decreases between $t=-0.001$ and $-0.2$ GeV$^2$ by about $\Delta B\approx 2$ GeV$^{-2},$ whereupon it stays constant until the dip, i.e. $|t|\approx1$ GeV$^2$.

At the LHC instead the "break" is exposed by TOTEM \cite{TOTEM8} by means of the normalized differential cross section  
\begin{equation} \label{Eq:norm}
R=\frac{\frac{d\sigma}{dt}-ref}{ref},
\end{equation}
where $ref=Ae^{Bt}$ with $A$ and $B$ determined from a fit to the experimental data. 

To be coherent with the TOTEM data, we show our calculations in the same way, i.e. by means of the normalized cross section Eq. \ref{Eq:norm}. With a simple Regge-pole model, shown in Sec.~\ref{s3} and used by us in Ref.~\cite{RPM}, we describe the low-$|t|$ LHC TOTEM 7 \cite{TOTEM7} and 8~TeV \cite{TOTEM8} data. The low-$|t|$ data can be well fitted also by a relevant form factor (residue function) in the scattering amplitude \cite{BPM} or an exponential with polynomial argument used by experimentalists \cite{TOTEM8}.  

\section{A simple Regge-pole model} \label{s3}
For our purposes we use a simple Regge pole model \cite{RPM} with a supercritical Pomeron $A_P$ and an effective Reggeon $A_f$ contributions:
\begin{equation} 
A(s,t)=A_P(s,t)+A_f(s,t),
\end{equation} 
where
\begin{eqnarray}
A_P(s,t)=-a_Pe^{b_P\alpha_P(t)}e^{-i\pi\alpha_P(t)/2}(s/s_0)^{\alpha_P(t)}, \nonumber \\ 
A_f(s,t)=-a_fe^{b_f\alpha_f(t)}e^{-i\pi\alpha_f(t)/2}(s/s_0)^{\alpha_f(t)}
\end{eqnarray}
with the trajectories
\begin{eqnarray}
\alpha_P(t)=\alpha_{0P}+\alpha'_Pt-\alpha_{1P}(\sqrt{4m_{\pi}^2-t}-2m_{\pi}), \nonumber \\  
\alpha_f(t)=\alpha_{0f}+\alpha'_ft-\alpha_{1f}(\sqrt{4m_{\pi}^2-t}-2m_{\pi}).
\end{eqnarray}
We use the norm:
\begin{equation}
\frac{d\sigma}{dt}(s,t)=\frac{\pi}{s^2}|A(s,t)|^2.
\end{equation}
Here $m_\pi$ is the pion mass, $4m_{\pi}^2=0.08$ GeV$^2$, $s_0=1$ GeV$^2$ and the free parameters are: $a_P$ ($\sqrt{mbGeV^2}$), $b_P$ (dimensionless), $\alpha_{0P}$ (dimensionless), $\alpha'_P$ (GeV$^{-2}$), $\alpha_{1P}$ (GeV$^{-1}$), $a_f$ ($\sqrt{mbGeV^2}$), $b_f$ (dimensionless), $\alpha_{0f}$ (dimensionless), $\alpha'_f$ (GeV$^{-2}$), $\alpha_{1f}$ (GeV$^{-1}$).

The result of fitting to the LHC TOTEM 7 and 8 TeV data are displayed in Sec.\ref{s4}. 
\section{Results of fits} \label{s4}
\begin{figure}[H] 
	\centering
	\includegraphics[width=.85\textwidth]{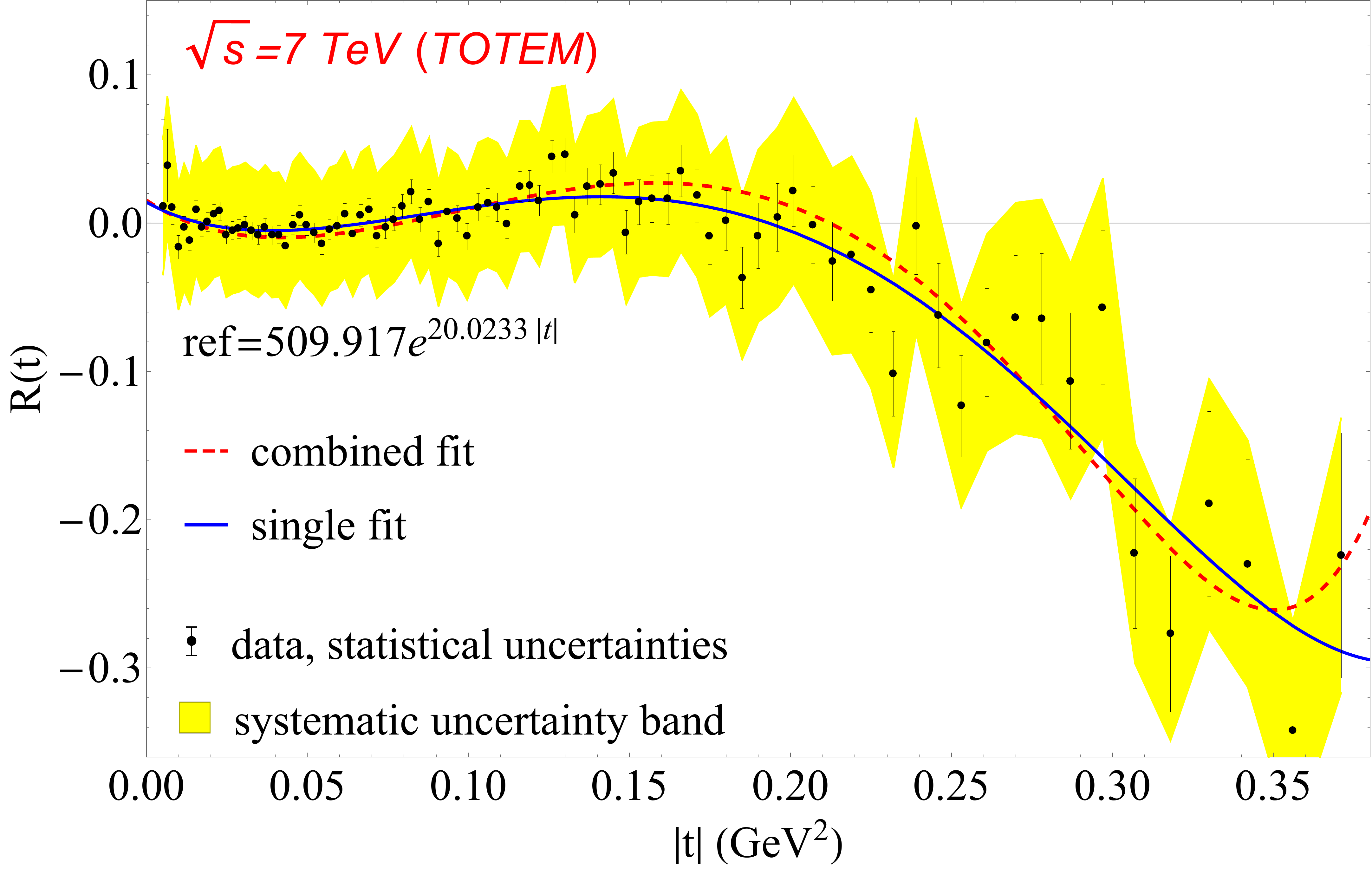}
	\caption{Normalized differential cross section calculated for TOTEM 7 TeV data.} 
	\label{Fig:7}
\end{figure}

\begin{figure}[H] 
	\centering
	\includegraphics[width=.85\textwidth]{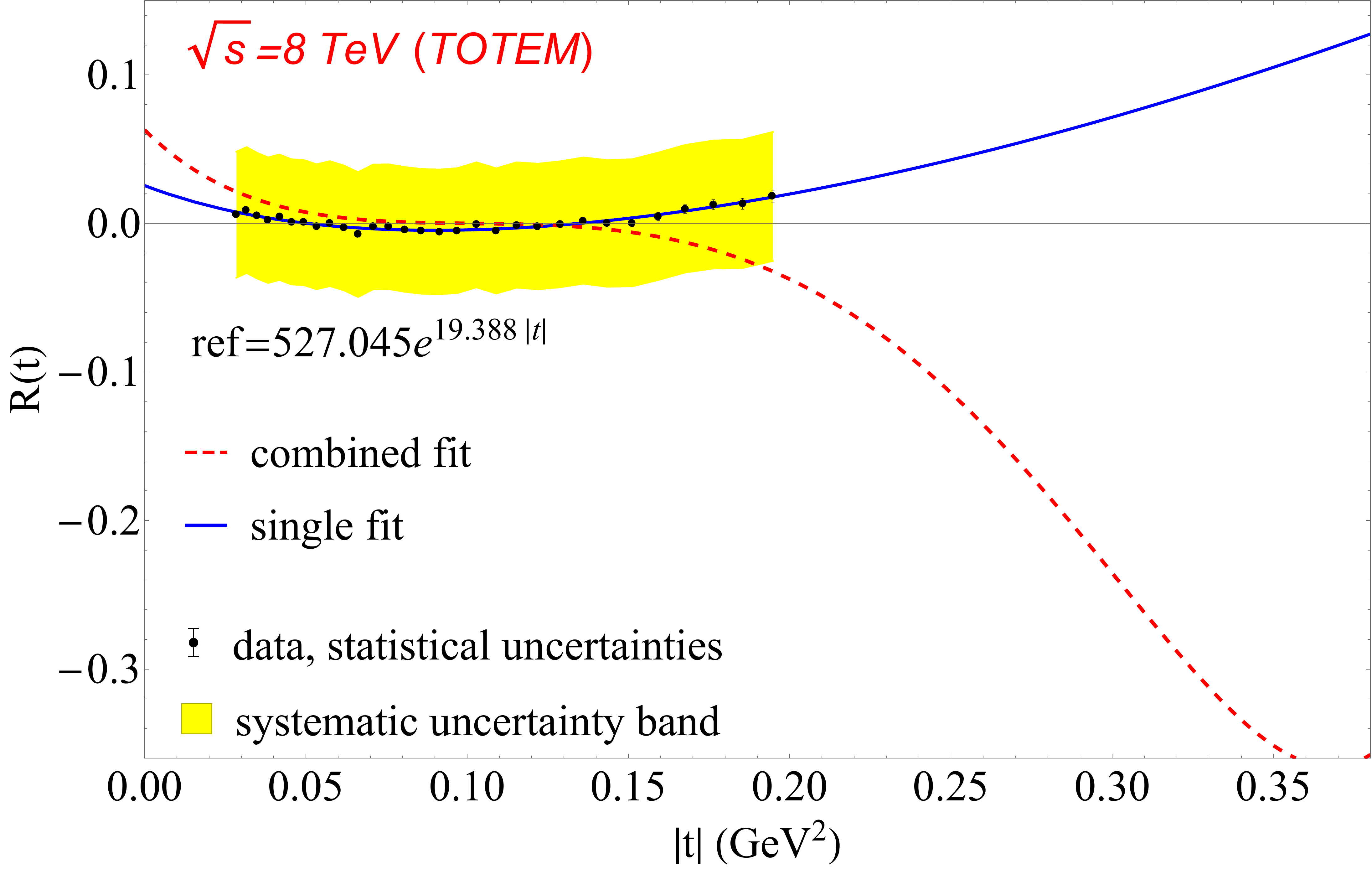}
	\caption{Normalized differential cross section calculated for TOTEM 8 TeV data.} 
	\label{Fig:8}
\end{figure}

Fitting the above described model to the LHC TOTEM data we take into account two cases: combined fit to the LHC TOTEM 7 and 8 TeV data; single fit to the two energy separately. The results of the fits are shown in Figs.~\ref{Fig:7} and \ref{Fig:8}. The values of fitted parameters are displayed in Table~\ref{ta1}.

Fig.~\ref{Fig:7} shows that the normalized cross section calculated for TOTEM 7 TeV data drastically deceases starting from $|t|\approx$~0.2 GeV$^2$. The reason of this phenomenon is the vicinity of the dip, acting as a "vortex" inclining differential cross section to drop towards the dip position. This behavior is not seen in Fig.~\ref{Fig:8} in the case of the TOTEM 8 TeV data, because the relevant $t$-range at this energy extends merely to $|t|$=0.2 GeV$^2$. However, the preliminary $13$ TeV data \cite{TOTEM13} beyond $|t|\approx$~0.2 GeV$^2$ show the decrease similarly that seen at 7 TeV.

\begin{table}[H]
	\centering
	{\begin{tabular}{@{}cccc@{}} \toprule
			Parameters & combined fit & single fit & single fit \\
			& (7 and 8 TeV)  & (only 7 TeV) & (only 8 TeV) \\ \colrule
			
			$\alpha_{0P}$ &1.08174 &1.0233&1.02768\\
			$\alpha'_P$   &0.128966&0.219225&0.451106\\
			$\alpha_{1P}$ &0.139175&0.159296&0.0378208\\
			$a_P$         &1.25014&1.8063&1.67174\\
			$b_P$         &1.26585&1.56255&1.52131\\ 
			$\alpha_{0f}$ &0.972729&0.707782&0.569488\\
			$\alpha'_f$   &0.227304&0.539584&0.266978\\
			$\alpha_{1f}$ &-0.0903918&-0.504727&0.0227016\\
			$a_f$         &-0.593097&-4.67554&-4.9071\\
			$b_f$         &3.20462&4.75995&3.71808\\ \botrule
			$dof$         &107&77&20\\
			$\chi^2/dof$  &0.14&0.15&0.18\\ \botrule
		\end{tabular} \label{ta1}}
	\caption{Values of fitted parameters in cases combined and single fits.}
\end{table} 

\section{Conclusions}
As emphasized at the beginning of the paper, the two structures on the otherwise exponential diffraction cone have quite different origin and physical meaning. 

We have used a simple Regge-pole model to interpolate between the low-$|t|$ elastic pp data measured by TOTEM collaboration. Although the combined fit to the 7 and 8 TeV data is not perfect, this model reproduces satisfactory the energy dependence as shown in Ref.~\cite{RPM}. 

Theoretical calculations of the relative contribution of the loop diagram, relative to the "Born term" (the ratio of two diagrams in the right-hand side of Fig.~\ref{Fig:Diagram}), is of great importance, however relevant calculations are beyond the capability of perturbative QCD. 

\section*{Acknowledgements}

We thank Jan Ka\v{s}par for an inspiring correspondence on the subject of this paper.

\end{document}